\documentclass[ 
aps,pra,twocolumn,
amsmath,amssymb,
,superscriptaddress]{revtex4}

\usepackage{graphicx}
\usepackage{dcolumn}
\usepackage{bm}
\usepackage{color}
\usepackage{amsmath} 
\usepackage{amssymb} 
\usepackage{amsxtra}
\usepackage{graphicx}
\usepackage{psfrag} 
\usepackage{bm} 
\usepackage{float} 
\usepackage{dsfont}
\usepackage{hyperref}
\input colordvi

\newcommand{\Cref}[1]{Chap.~\ref{#1}}

\newcommand{\fref}[1]{Fig.~\ref{#1}}

\accentedsymbol{\dbareps}{\Bar{\Bar{\epsilon}}}
\accentedsymbol{\dbarA}{\Bar{\Bar{A}}}
\accentedsymbol{\dbarB}{\Bar{\Bar{B}}}

\begin{document}

\title{Effects of Random Link Removal on the Photonic Band Gaps of Honeycomb Networks}

\author{Marian Florescu} 

\email{florescu@princeton.edu}
\affiliation{Department of Physics, Princeton University, Princeton,  New Jersey 0855,  USA}
\author{Salvatore Torquato} 
\affiliation{Department of Chemistry, Princeton University, Princeton,  New Jersey 0855,
  USA}
\affiliation{Department of Physics, Princeton University, Princeton,  New Jersey 0855,  USA}
\affiliation{Department of Chemistry, Princeton University, Princeton,  New Jersey 0855,
  USA}
\affiliation{Princeton Center for Theoretical Science, Princeton University, Princeton,
  New Jersey 0855, USA}

\affiliation{Princeton Institute for the Science and Technology of Materials, Princeton University, Princeton, New Jersey 08544, USA}
\author{Paul J. Steinhardt}
\affiliation{Department of Physics, Princeton University, Princeton,  New Jersey 0855,  USA}
\affiliation{Princeton Center for Theoretical Sciences, Princeton University, Princeton,  New Jersey 0855, USA}

\date{\today}
\begin{abstract}

We explore the effects of random link removal on the photonic band gaps of honeycomb
networks. Missing or incomplete links are expected to be common in practical realizations
of this class of connected network structures due to unavoidable flaws in the fabrication
process. We focus on the collapse of the photonic band gap due to the defects induced by
the link removal. We show that the photonic band gap is quite robust against this type of
random decimation and survives even when almost 58\% of the network links are removed.

\end{abstract}

\maketitle

Photonic crystals, and in particular photonic band gap (PBG)
materials,\cite{john87,yablonovitch87} constitute a class of dielectric materials, in
which the basic electromagnetic interaction is controllably altered over certain
frequencies and length scales.\cite{ref_3,ref_4} The ability to tailor the photonic
density of states in a prescribed manner and to engineer the symmetry properties of the
electromagnetic field inside a PBG material enables the design of materials and devices
that control the flow and the emission and absorption properties of electromagnetic
radiation.\cite{ref_5,ref_6,ref_7,ref_8} Most of the photonic crystals fabricated to date
possess intrinsic disorder, including size variation among the scattering centers,
roughness of the dielectric walls, point defects, lattice dislocation and stacking
defaults.\cite{ref_9,ref_10} The presence of disorder results in defects modes that can
fill the PBG region partially or completely and alter significantly the photonic
properties of the dielectric crystals.

\begin{figure}[ht]
\centerline{ \includegraphics[width=1\linewidth]{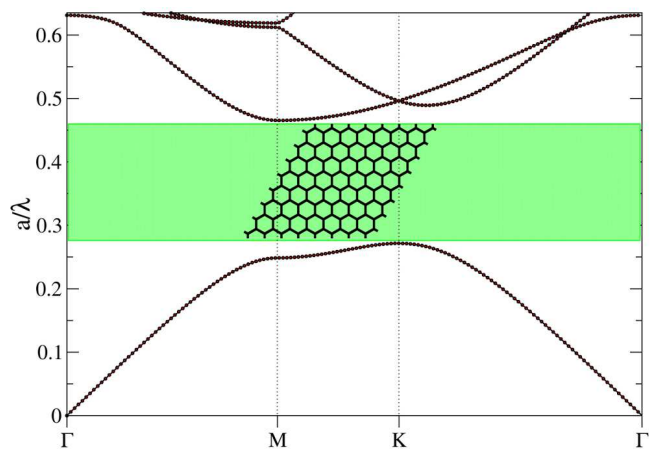} }
\caption{\label{fig1} (Color online) Band structure for the optimal unperturbed TE PBG
  structure consisting of dielectric walls with dielectric constant $\epsilon=11.56$
  thickness $w/a=0.94$ (shown as an inset), which presents a TE PBG of $\Delta
  \omega/\omega_C=52.63\%$, the largest known PBG in two-dimensional photonic crystals for
  this dielectric constant contrast ratio.}
\end{figure}

In this paper we focus on the effect of random link removal on the properties of photonic
gaps of honeycomb networks.  It is well established that two-dimensional dielectric
structures based on connected network architectures favor the formation of large PBGs for
transverse electric (TE) polarized radiation. \cite{ref_11,ref_12,ref_13} The largest
known TE PBG is achieved in a dielectric structure obtained by decorating the links of a
honeycomb network lattice with dielectric walls of a certain thickness \cite{ref_13_a}
(also depicted in the inset of \fref{fig1}).

We consider a special type of disorder produced by the random removal of links from a
honeycomb connected network. This type of disorder is expected to be common in practical
realizations of this class of connected network structures due to unavoidable flaws in the
fabrication process.  The major goals of this study are to explore the robustness of the
PBG against random removal of links and to establish a relationship between the topology
of defect complexes (resulting from the removal of multiple links) and the size of the
band gap.
\begin{figure}[ht]
{\centerline{ \includegraphics[width=1\linewidth]{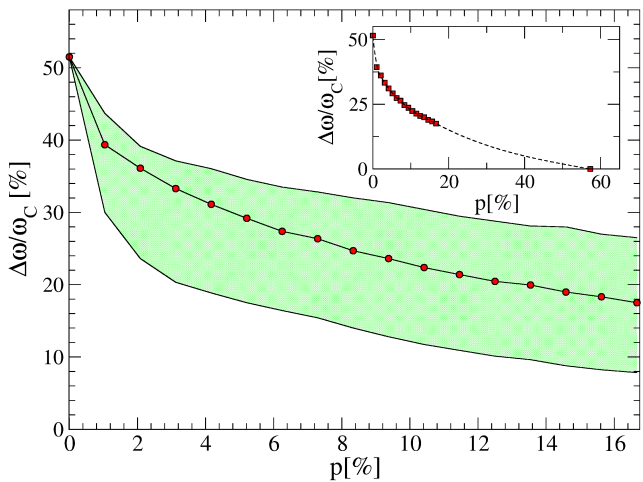}}}
\caption{\label{fig2} (Color online) The small filled circles represent the PBG size as a
  function of the percentage of randomly removed links, as obtained by averaging over a
  statistical ensemble of 1500 configurations. The upper and lower lines represent the
  maximum and minimum value among the ensemble. The inset shows the extrapolation of the
  average PBG to percentages large enough to close the photonic band gap.} 
\end{figure}
The class of photonic structures considered in this study are generated by decorating the
sides of a two-dimensional honeycomb tessellation with dielectric walls of dielectric
constant $\epsilon=11.56$ (corresponding to Si in the infrared region) and thickness
$w/a=0.94$, where $a$ is the honeycomb lattice constant, corresponding to the largest
known TE PBG: $52.63\%$ (see \fref{fig1}).

To evaluate the influence of the link disorder on the band gap properties, we construct a
finite domain of the honeycomb dielectric network and define a statistically independent
probability, $p$, for the removal of the dielectric links (i.e., Bernoulli percolation
\cite{ref_14_a, ref_14}) that connect the vertices of the honeycomb network. For each $p$,
a statistical ensemble consisting of 1500 configurations is generated, and the photonic
band structure is calculated along a closed contour around the Brillouin zone (as in
Figs. 1, 3a and 4a) for each member of the ensemble.  In addition, we have checked that
gap size by calculating the band structure for 100 ${\bf k}$-points randomly distributed
in the first Brillouin zone. Due to computational limitations, the domain considered
consists of $8\times 8 $ unit cells.
\begin{figure}[ht]
\centerline{ \includegraphics[width=\linewidth]{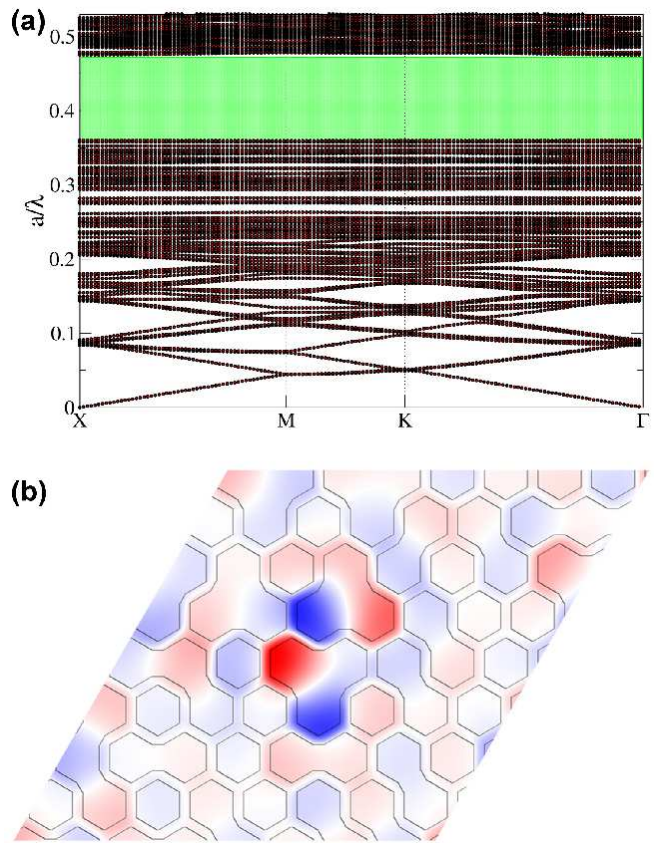}}
\caption{\label{fig3} (Color online) The connected network photonic structure with the
  largest TE PBG for $p=16.7\%$ (selected from 1500 randomly generated configurations). a)
  The photonic band structure for dielectric constant contrast $\epsilon=11.56$ and wall
  thickness $w/a=0.94$, displaying a PBG with $\Delta\omega/\omega_C=27.78\%$ (shaded
  region).  b) The magnetic field distribution for the mode at the lower edge of the PBG
  (the black lines show the contour of dielectric material distribution).}
\end{figure}
In \fref{fig2}, we plot the maximum, minimum and average PBG in a given ensemble of
randomly generated structures as a function of the number of links removed. We note that
while full connectivity is associated with the largest PBG structures \cite{fan}, the band
gap is very resilient against line removal; even after the random removal of $p=16.7\%$ of
all the links, the structures present a sizable average band gap of
$\Delta\omega/\omega_C=17\%$ (here $\Delta\omega$ is the PBG size and $\omega_C$ the gap
central frequency). Moreover, the minimal PBG configuration still retains a gap of about
8\%, while the maximal PBG configurations display sizable band gaps of about 28\%.
\begin{figure}[ht]
\centerline{\includegraphics[width=\linewidth]{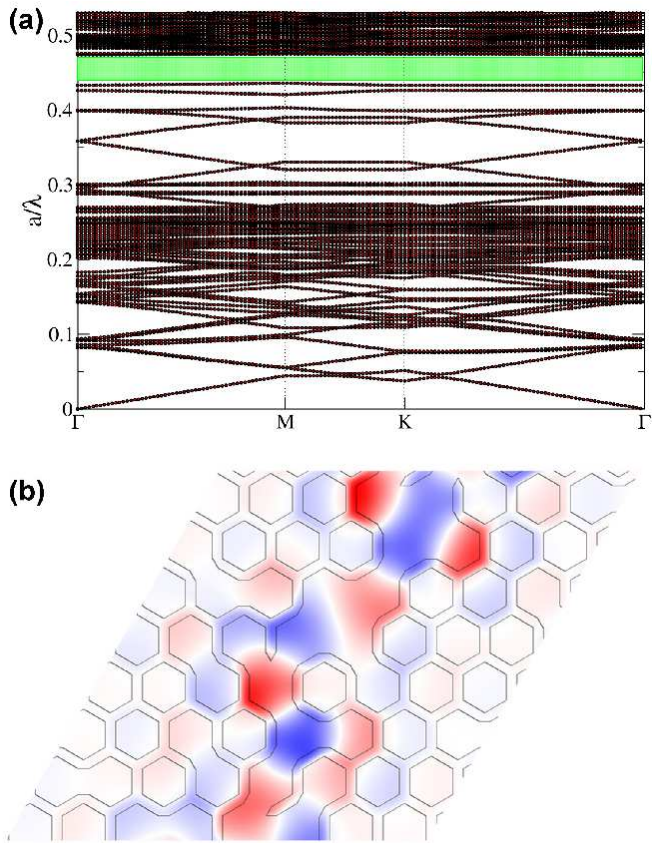}}
\caption{\label{fig4} (Color online) The connected network photonic structure presenting
  the smallest TE PBG for $p=16.7\%$ (selected from 1500 randomly generated
  configurations). a) The photonic band structure for dielectric constant contrast
  $\epsilon=11.56$ and wall thickness $w/a=0.94$, displaying a PBG
  $\Delta\omega/\omega_C=7.95\%$.  b) The magnetic field distribution for the mode
  corresponding to the lower edge of the PBG (the black lines show the contour of
  dielectric material distribution).}
\end{figure}

We have analyzed different types of defects generated by the link removal and their
influence on the band gap. In general, the link removal results in a lowering of the local
dielectric constant, which, as expected, generates an increase in the frequency of the
photonic modes around the lower edge of the PBG. These modes ascend inside the PBG and
subsequently the PBG size is reduced. As the density of the removed links increases, the
unoccupied links may cluster together to generate larger defect complexes. Isolated
clusters containing a small number of missing links generate cavity-like
defects. Cavity-like defects induce photonic modes with a high degree of localization and
low dispersion, and generate defect bands with small spectral bandwidth.

In \fref{fig3} a we plot the band structure for the configuration that displays the
largest gap ($\Delta\omega/\omega_C=27.78\%$) in a structure with $p=16.7\%$.  We note
that the modes shift to inside the PBG due to the presence of material defects present
minimal dispersion. The corresponding dielectric structure (the black-line contour in
\fref{fig3} b) and the magnetic field distribution of the photonic mode defining the lower
edge of the PBG confirm that most of the defect modes are confined around the random
cavity-like defects in the structure.

A second type of defect is a waveguide-like complex, which occurs when removed links
concatenate to create a defect-cluster spanning the entire structure. In this case, the
defect modes are associated with propagation along the waveguide channel, thus localized
only transverse to the waveguide. As a result, they display stronger dispersion and have a
more deleterious effect on the width of the PBG.

In \fref{fig4} a, we plot the band structure for the configuration that achieves the
minimal PBG ($\Delta\omega/\omega_C=7.96\%$) among structures with $p=16.7\%$. Here we
note that among the modes that partially fill the PBG region, there are modes showing
considerable dispersion.  As shown in \fref{fig4} b, the unoccupied links form a waveguide
channel (the dielectric profile is shown by the black-line contour), and the mode at the
lower edge of the PBG has the magnetic field rather tightly confined inside the waveguide.

To summarize, we have analyzed the effects of link disorder on the TE band gaps in
honeycomb connected network photonic structures. We have demonstrated that the PBG is
robust against this type of disorder and have identified the relevant defect topologies.
Removing links of dielectric material introduces zero-dimensional (cavities) or
one-dimensional (waveguide channels) defects, and, depending on their spectral location
and dispersion, the PBG size is reduced as $p$ increases. At some threshold value of $p$,
the PBG vanishes.  Thus, our study can be regarded to be a {\it photonic percolation}
problem.  We define the {\it photonic percolation threshold} as the value $p=p_c$ at which
the PBG first vanishes in the infinite-system limit. For $p \ge p_c$, the PBG is
identically zero.  Computational requirements make difficult a precise determination of
the {\it photonic percolation threshold}, but we have performed an extrapolation of our
results (shown in the inset of \fref{fig2}), which indicates that the PBG vanishes if, on
average, $p_c=57.6\%$ of the links are removed. Further band structure calculations
confirm that, if $57.6\%$ of the connected network links are randomly removed, the average
PBG collapses. This photonic-percolation-threshold value should be contrasted with that of
Bernoulli nearest-neighbor bond percolation on a honeycomb network for which
$p_c=1-2\sin\pi/18\approx 65.28\%$. The fact that the values of $p_c$ are not identical is
not surprising because they depend in detail on different physical characteristics.  Other
percolation phenomena with yet different values $p_c$ include elasticity percolation
\cite{ref_14,ref_15} and quantum percolation.~\cite{ref_16}

This work was supported by the MRSEC Program of the National Science Foundation under
Awards DMR-0820341 and ECCS-1041083.

\end{document}